\documentclass[aps,prl,twocolumn,amsmath,amssymb,floatfix]{revtex4}
\usepackage{tabularx}
\usepackage{bm}
\usepackage{euscript}
\usepackage{epsfig,psfrag,subfigure}
\usepackage{graphicx}
\usepackage{color}
\usepackage{amsfonts}
\usepackage{exscale}
\usepackage{amsbsy}
\usepackage{dsfont}

\newcommand{\br}{\mathbf{r}}

\newcommand{\bk}{\mathbf{k}}

\newcommand{\bq}{\mathbf{q}}
\newcommand{\bQ}{\mathbf{Q}}

\newcommand{\eff}{\mathrm{eff}}

\def\prl#1#2#3{Phys.\ Rev.\ Lett.\ {\bf #1}, #2 (#3)}

\def\prb#1#2#3{Phys.\ Rev.\ B {\bf #1}, #2 (#3)}

\def\rmp#1#2#3{Rev.\ Mod.\ Phys.\ {\bf #1}, #2 (#3)}

\begin{document}
\title{Algebraic spin liquid in an exactly solvable spin model}
\author{Hong Yao, Shou-Cheng Zhang, and Steven A. Kivelson}
\affiliation{Department of Physics, Stanford University, Stanford, California 94305}
\date{\today}
\begin{abstract}
We have proposed an exactly solvable quantum spin-3/2 model on a square lattice. Its ground state is a quantum spin liquid with a half integer spin per unit cell. The fermionic  
excitations are gapless with a linear dispersion, while the topological ``vison'' excitations are gapped. Moreover, the massless Dirac fermions are stable. Thus, this model is, to the best of  our knowledge, the first exactly solvable model of half-integer spins whose ground state is an ``algebraic spin liquid.''  
\end{abstract}
\maketitle
The term ``spin liquid'' is widely used to give sharp meaning to the more general intuitive notion of a Mott insulator;  a spin liquid is an insulating state that cannot be adiabatically connected to a band insulator, {\it i.e.} to an insulating Slater determinant state.   In a system that preserves time reversal symmetry, any insulating state with an odd number of electrons (or a half-integer spin) per unit cell is  
a spin liquid.  Interesting proposals \cite{Anderson87,wenleereview}  have been made concerning the relevance of such states to the theory of high temperature superconductivity in the cuprates and other materials.  
Indeed, various spin-liquid phases have been proposed, which are distinguished by  
the character of any gapless spinons and the exchange statistics of the topological ``vison'' excitations. 

Since they are new and ``exotic'' quantum phases of matter, 
it is desirable to construct solvable models with short range interactions with stable spin-liquid ground-state phases. 
A breakthrough occurred when Moessner and Sondhi \cite{Moessner01} demonstrated the existence of a 
gapped spin liquid ground state in the quantum dimer model \cite{Rokhsar88}, analogous to the short range version  of the RVB state \cite{Anderson73,Kivelson87}.  An exactly solvable spin-1/2 model in a gapped $Z_2$ spin liquid phase was later constructed by Wen \cite{Wen03}. However, much of the recent interest, spurred in part by the possible observation of such a state in  $\kappa$-(ET)$_2$Cu$_2$(CN)$_3$ \cite{Shimizu03,Yamashita08,Sachdev08} and Zn(Cu)$_3$(OH)$_6$Cl$_2$ \cite{Helton07}, has focussed on spin-liquids with gapless spin excitations, so-called ``algebraic spin liquids.''

The exactly solvable Kitaev model on the honeycomb lattice \cite{Kitaev06} can  
exhibit gapless excitations. However, because the honeycomb lattice has two sites per unit cell, this model has an integer spin, hence an even number of electrons per unit cell. 
In the present paper we construct an exactly solvable model, in much the same spirit as the Kitaev model, whose ground state is a spin liquid with an {\it odd} number electrons per unit cell and stable gapless fermionic  
excitations. To the best of our knowledge, this is the first exactly solvable model with this sort of spin liquid ground state -algebraic spin liquid.

The Kitaev model has a spin-1/2 on each site of a trivalent lattice, where the coordination number is dictated by the existence of  three Pauli matrices. In order to study a model on a square lattice, we instead consider a model with a spin-3/2 on each lattice site.  The resulting larger Hilbert space, with 4 spin polarizations per site, permits us to express the model in terms of the $4 \times 4$ anticommuting Gamma matrices, $\Gamma^a$ ($a=1,\cdots,5$) which form Clifford algebra, $\{\Gamma^a,\Gamma^b\}=2\delta^{ab}$. Specifically,  the 5 Gamma matrices can be represented \cite{Zhang04} by symmetric bilinear combinations of the components of a spin 3/2 operator, $S^\alpha$, as 
\begin{eqnarray}
&&\!\!\!\!\!\!\Gamma^1=\frac{1}{\sqrt{3}}\{S^y,S^z\}, \Gamma^2=\frac{1}{\sqrt{3}}\{S^z,S^x\},
\Gamma^3=\frac{1}{\sqrt{3}}\{S^x,S^y\},\nonumber\\
&&\qquad\Gamma^4=\frac{1}{\sqrt{3}}[(S^x)^2-(S^y)^2],
~\Gamma^5=(S^z)^2-\frac{5}{4}.
\label{Eq:Gamma}
\end{eqnarray}

{\it Model Hamiltonian}:
We define our model on a square lattice, with a spin-3/2 on each site, and corresponding $\Gamma$ matrices defined as in Eq. (\ref{Eq:Gamma}). In terms of these, 
\begin{eqnarray}
&&{\cal H}=\sum_{i}\Big[J_x \Gamma^1_i \Gamma^2_{i+\hat x} +J_y \Gamma^3_i\Gamma^4_{i+\hat y}\Big]\\
&&~~~+\sum_i\Big[J_x^\prime \Gamma^{15}_i\Gamma^{25}_{i+\hat x}+J_y^\prime \Gamma^{35}_i\Gamma^{45}_{i+\hat y}\Big]
-J_5\sum_i \Gamma^5_i,\nonumber
\end{eqnarray}  
where $\Gamma^{ab}\equiv[\Gamma^a,\Gamma^b]/(2i)$  
and $i$ labels the lattice site at $\br_i=(x_i,y_i)$. We call this model the Gamma matrix model (GMM). Suppose that the square lattice has $N=L_xL_y$ sites, where $L_x$ and $L_y$ are the lattice's linear sizes and are assumed, for simplicity, to be even in this paper. Moreover, we consider periodic boundary conditions. Obviously, the GMM can be written explicitly as a spin-3/2 model. The GMM model respects translational symmetry and time reversal symmetry (TRS).  
It does not have  
global spin SU(2) or even U(1) rotational symmetry, but 
is invariant under 180$^\circ$ rotations about the $z$ axis in spin space, {\it i.e.} it has global  
Ising symmetry.  
The lack of any continuous symmetry implies that the fermionic excitations (discussed below) do not have well-defined spin quantum number, so their relation to ``spinons'' is unclear. 
A feature of the  
model which makes it solvable is 
 an infinite set of conserved ``fluxes:'' $[\hat W_i, {\cal H}]=0$ for any $i$ and $[\hat W_i,\hat W_j]= 0$,   
 where $\hat W_i \equiv\Gamma^{13}_i \Gamma^{23}_{i+\hat x} \Gamma^{14}_{i+\hat y} \Gamma^{24}_{i+\hat x+\hat y}$ is the plaquette operator at site $i$.

Note that, in contrast to the spin-1/2 Kitaev model on the honeycomb lattice,  the GMM has an odd number (3) of electrons per unit cell.  The present model in the  
limit $J_x=J_y=J'_x=J'_y$ and $J_5=0$ is similar to a model proposed by Wen in Ref. \cite{Wen03PRD}, where, however, the Gamma matrices were constructed from two spin-1/2 operators on each site, and therefore behave differently under time reversal than in the present realization. Moreover, the present model generically does not possess the special symmetries  
of Wen's model that are responsible for some of the behaviors of that model.

{\it Fermionic Representation}: 
Spin-3/2 operators can be expressed as bilinear forms involving three flavors of fermion operators, $S^z =a^\dagger a + 2 b^\dagger b - 3/2$ and $S^+=\sqrt{3}f^\dag a+\sqrt{3} a f+2a^\dag b$, subject to the constraint that the  physical states are only those with odd ``fermion parity,'' $(-1)^{\hat N}=-1$, where $\hat N = f^\dagger f +a^\dagger a+b^\dagger b$. Rather than using this representation in terms of three Dirac fermions, we will directly represent the Gamma matrices in term of a related set of 6 Majorana fermions: 
\begin{eqnarray}
\Gamma^\mu_i=ic^\mu_i d_i,~\Gamma^{\mu 5}_i=ic^\mu_i d'_i, ~\mu=1,2,3,4,~\Gamma^5_i=id_id'_i
\end{eqnarray} 
where $c^\mu_i$, $d_i$, and $d'_i$ are Majorana fermions on site $i$. Six Majorana fermions form an eight dimensional Hilbert space, which is an enlarged one from the physical Hilbert space of an spin-3/2.  
In terms of spin-3/2 operators, $\Gamma^1_i\Gamma^2_i\Gamma^3_i\Gamma^4_i\Gamma^5_i=-1$ for all $i$. Consequently, all allowed physical states $|\Psi\rangle$ in term of Majorana fermions must satisfy the following constraint, for all $i$,
\begin{eqnarray}
D_i|\Psi\rangle\equiv\left[-ic^1_ic^2_ic^3_ic^4_id_id'_i\right] |\Psi\rangle =|\Psi\rangle.
\end{eqnarray}

In terms of Majorana fermions, the Hamiltonian in the enlarged Hilbert space can be written as
\begin{eqnarray}\label{H_MF}
&&{\cal H}=\sum_i\Big[J_x \hat u_{i,x}id_id_{i+\hat x} +J_y \hat u_{i,y}id_id_{i+\hat y}
\nonumber\\
&&+J_x^\prime\hat u_{i,x}id'_id'_{i+\hat x} +J^\prime_y \hat u_{i,y}id'_id'_{i+\hat y} 
- J_5 id_i d'_i\Big],
\end{eqnarray}
where $\hat u_{i, x}\equiv-ic^1_ic^2_{i+\hat x}$ and $\hat u_{i,y}\equiv-ic^3_ic^4_{i+\hat y}$. 
It is obvious that $\hat u_{i,\lambda}$ are conserved quantities with eigenvalues $u_{i,\lambda}=\pm 1$, $\lambda=x,y$. The enlarged Hilbert space can be divided into sectors $\{u\}$. In each sector, the Hamiltonian Eq. (\ref{H_MF}) describes free Majorana fermions:
\begin{eqnarray}\label{H_MF2}
&&{\cal H}(\{u\})=\sum_i \Big[J_x u_{i,x}id_id_{i+\hat x} +J_y u_{i,y}id_id_{i+\hat y}
\nonumber\\
&&\qquad+J_x^\prime u_{i,x}id^\prime_id^\prime_{i+\hat x} +J^\prime_y u_{i,y}id^\prime_id^\prime_{i+\hat y}-J_5id_id'_i \Big],
\end{eqnarray}
where $u_{i,\lambda}$ are emergent $Z_2$ gauge fields.  
The $Z_2$ gauge transformations are given by $d_i\to \Lambda_i d_i$, $d'_i\to \Lambda_i d'_i$, and $u_{i,\lambda}\to \Lambda_i u_{i,\lambda} \Lambda_{i+\lambda}$, 
where $\Lambda_i =\pm 1$. In the enlarged Hilbert space, the 
eigenstates $|\psi\rangle=|\psi\rangle_c\otimes|\psi\rangle_{d,d'}$ 
can be written as a direct product of a part that involves the $c^\mu$ fermions and a part that 
involves the 
$d$  and $d'$ fermions, respectively.
Here $|\psi\rangle_c$ is defined by $\hat u_{i,\lambda}|\psi\rangle_c = u_{i,\lambda} |\psi\rangle_c$
and $|\psi\rangle_{d,d'}$ are eigenstates of Eq. (\ref{H_MF2}) with the corresponding $u_{i,\lambda}$'s. 

The spectrum of ${\cal H}(\{u\})$ depends only on gauge invariant quantities - the flux on local plaquettes, 
$\exp(i\phi_i)\equiv u_{i,x}u_{i+\hat x,y}u_{i,y} u_{i+\hat y,x}$, and two global fluxes $\exp(i\phi_x)\equiv \prod_{i(y_i=1)} u_{i,x}$ and $\exp(i\phi_y)\equiv \prod_{i(x_i=1)} u_{i,y}$, 
where $\phi_i$ and $\phi_{x,y}=0,\pi$. 
[Note that the previously defined $W_i=-\exp(i\phi_i)$.] It is obvious that $\sum_i \phi_i=0$ (mod $2\pi$), so there are $N-1$ independent local fluxes. Including the two global fluxes, the number of independent fluxes is $N+1$. Since there are $2N$ $Z_2$ gauge fields, the number of different gauge field choices corresponding to each flux sector $\{\phi\}$ is $2^{N-1}$. In other words, 
in the enlarged Hilbert space, each state is 
$2^{N-1}$-fold degenerate. Note that in the thermodynamic limit, the energy is independent of the  two global fluxes, which gives rise to a fourfold topological degeneracy of the physical ground states. 

{\it Projection operators}:
Most of the states in the enlarged Hilbert space are not physical states. To obtain a physical eigenstate, we must find a linear combination of the degenerate eigenstates which is simultaneously an eigenstate of every $D_i$ with eigenvalue $1$. This is realized  
through the projection operator $P$:
\begin{eqnarray}
|\Psi\rangle= P|\psi\rangle \equiv \prod_{i}\Big[(1+D_i)/2\Big]|\psi\rangle.
\label{Eq:proj}
\end{eqnarray}
Clearly, Eq. (\ref{Eq:proj}) 
implies $D_i|\Psi\rangle=|\Psi\rangle$ for any $i$.  
Explicitly, $P$ is given by
\begin{eqnarray}
P=\Big[1+\sum_i D_i+\sum_{i_1<i_2}D_{i_1}D_{i_2}+\cdots+\prod_i D_i\Big]/2^N.
\label{Eq:P}
\end{eqnarray}  
$D_i$ acting on a direct product state $|\psi\rangle$ is equivalent to a gauge transformation on site $i$.  A subtlety here is that there are $2^N$ operators in the sum in Eq. (\ref{Eq:P}), but only $2^{N-1}$ inequivalent gauge transformations. In fact, $D\equiv \prod_i D_i$ implements a gauge transformation on every site, thus leaving all gauge fields unchanged.  
It follows that $P=P'(1+D)$, where $P'$ includes all inequivalent transformations. Since $[D,{\cal H}]=0$ and $D^2=1$, $D|\psi\rangle=\pm|\psi\rangle$.  
Moreover, $D=\prod_i\big[\hat u_{i,x}\hat u_{i,y}\big] \prod_i\big[id_id'_i\big]$. In terms of the number $\hat N_f$ of the Dirac fermions,  
\begin{equation}
f_j\equiv i^j(d_j+id_j')/2, ~\hat N_f=\sum_i f^\dag_i f_i
\end{equation}
and fluxes, $\hat N_\phi$, we obtain $D=(-1)^{\hat N_\phi}(-1)^{\hat N_f}$.
$\hat N_f$  
is conserved modulo 2 by the Hamiltonian. $\hat N_\phi$ is defined by dividing the plaquettes into two sublattices and counting the number of $\pi$-fluxes through one or the other sublattice \footnote{The total number of $\pi$-fluxes must be even,  
so the choice of which sublattice is immaterial.}.

Thus, depending on the fermion and flux parity, $P$ either annihilates a given direct product state, $|\psi\rangle$ , or maps it to the equal weight linear superposition of all gauge transformations acting on $|\psi\rangle$.
For instance, when there is a $\pi$-flux through each plaquette, which, as discussed below, is the ground state sector, $D=(-1)^{\hat N_f}$, {\it i.e.} all physical states must have an even number of fermions. Conversely, where 1 $\pi$-flux is added to each sublattice, only states with $\hat N_f = $ odd survive projection. Conserving the parity of fermion number reflects the fact 
that physical fermionic excitations are created by non-local (string)  
 operators \cite{Wen03PRD,HY}. 

{\it $\pi$-flux state and gapped visons}:
In each flux sector $\{\phi\}$, the lowest energy of the Hamiltonian is denoted by $E_0(\{\phi\})$. The ground state energy of the model is achieved by minimizing $E_0(\{\phi\})$ with respect to $\{\phi\}$. Formally, by integrating out the fermions, an effective action for the $Z_2$ gauge fields can be derived. However, in general, it is nontrivial to obtain an explicit form of the effective action.  

For $J_5=0$, fortunately, there is a theorem due to Lieb \cite{Lieb94} which implies  that the  
energy minimizing flux sector of a half-filled band of electrons hopping on a planar, bipartite lattice is $\pi$ per square plaquette. 
We define this uniform $\pi$-flux, the ground state sector, as being vortex-free. $Z_2$ vortex excitations, ``visons'', are defined to be plaquettes with $\phi_i=0$. Due to the constraint $\sum_i \phi_i=0$ (mod $2\pi$), only an even number of visons are allowed. It is special for this model that visons do not have dynamics even though they interact with one another. (Numerical calculations of the two vison energy reveals that the interaction between visons is repulsive.) The minimum energy cost of creating two visons in the ground state is always finite and is defined as the vison gap energy $\Delta_v\sim(\sqrt{|J_xJ_y|}+\sqrt{|J'_xJ'_y|})$.  
Since the visons are gapped, the uniform $\pi$-flux is still the ground state sector as long as $J_5$ is much smaller than  $\Delta_v$. In the rest of paper, we restrict our attention to cases (e.g. small $J_5$) in which the ground state lie in the $\pi$-flux sector. 

{\it Spin correlation}:
The state with uniform $\pi$-fluxes preserves the translational symmetry and TRS of the model. 
To prove the ground state of the model is a spin liquid, we need to show that there is no magnetic order and that the spin correlations are short-ranged. Spin-3/2 operators can be expressed as bilinear forms of Majorana fermions. For instance, $S^z_i=ic^3_ic^4_i+\frac{1}{2}ic^1_ic^2_i$. The action of $S^z_i$ on the ground state $|\Psi\rangle$ creates visons in the four surrounding plaquettes around site $i$ \cite{Baskaran07}. The effect of $S^x_i$ and $S^y_i$ on the ground state 
is similar. Because visons are non-dynamical 
in the present model, $\langle \Psi| S^\alpha_i S^\beta_j |\Psi\rangle$ is identically zero unless sites 
$i$ and $j$ are nearest neighbors, i.e. the spin correlations are unphysically short-ranged. 
Presumably, if additional small, local terms are added to the Hamiltonian, 
perturbative 
corrections to this correlator  
would lead to a finite correlation length.  

{\it Gapless fermions}: 
To obtain the excitation spectrum in the $\pi$-flux (ground state) sector, we fix the gauge by choosing $u_{i,x}=1,u_{i,y}=(-1)^i$. 
The corresponding Hamiltonian is given by  
\begin{eqnarray}\label{pwave}
&H_0=\sum_i\Big[t_xf^\dag_if_{i+\hat x}+t_y(-1)^if^\dag_if_{i+\hat y}-J_5f^\dag_i f_i\nonumber\\
&\qquad\qquad-\Delta_x(-1)^if^\dag_if^\dag_{i+\hat x} -\Delta_y f^\dag_i f^\dag_{i+\hat y}+H.c.\Big],
\end{eqnarray}
which describes a $p$-wave superconductor of spinless fermions \cite{Chen07}. Here $t_\lambda\equiv J_\lambda+J'_\lambda$ and $\Delta_\lambda\equiv J_\lambda-J'_\lambda$. (Note that in Ref. \cite{Wen03PRD}, the pairing terms are absent due to  ``projective symmetries''.) In terms of the Bloch states, $f_{\bk} =\sum_i \exp(-i\bk\cdot\br_i) f_i/\sqrt{N}$, 
Eq. (\ref{pwave}) in momentum space is given by
\begin{eqnarray}
H_0=\sum_\bk \Phi^\dag_\bk H_\bk \Phi_\bk,~\Phi^\dag_\bk=(f^\dag_\bk, f^\dag_{\bk+\bQ} ,f_{-\bk} ,f_{-\bk-\bQ}),
\end{eqnarray}
where $\bQ=(\pi,\pi)$ and the summation is over only half of the Brillouin zone since $\bk$ is equivalent to $\bk+\bQ$. In general, the analytical form of the eigenvalues of the $4\times 4$ matrix $H_\bk$ are complicated. Due to time reversal symmetry, however, it is straightforward to derive the quasiparticle excitation spectrum as follows: 
\begin{eqnarray}
E_{\pm,\bk}=2\sqrt{J_5^2+2g_{+,\bk} \pm 2\sqrt{g^2_{-,\bk}+J_5^2{g_\bk}}},
\end{eqnarray}
where $g_{\pm,\bk}=(J_x^2\pm {J'_x}^2)\cos^2 k_x +(J_y^2\pm {J'_y}^2)\sin^2 k_y$ and $g_\bk=(J_x+J'_x)^2\cos^2 k_x +(J_y+J'_y)^2\sin^2 k_y$. 

When $J_5=0$, 
$E_{+,\bk} =4({J_x^2\cos^2 k_x+J_y^2\sin^2 k_y})^{1/2}$ and $E_{-,\bk}=4({{J'_x}^2\cos^2 k_x+{J'_y}^2\sin^2 k_y})^{1/2}$  
are the energies of the $d$ and $d'$ Majorana fermions respectively since they decouple from each other. Both spectra are gapless at nodes $\pm K=\pm(\pi/2,0)$, around which the spectrum is linear in momentum; the excitations are massless Dirac fermions. However, the massless fermions 
 are unstable in the sense that additional small, local (further neighbor hopping and pairing) terms can gap the 
 nodes \cite{HY}. 

\begin{figure}[t]
\includegraphics[scale=0.7]{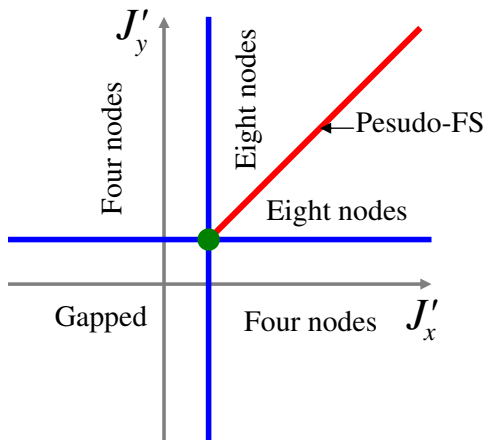}
\caption{The quantum phase diagram of the Gamma matrix model as a function of $J'_x$ and $J'_y$ in the case $J_5\ll J_x, J_y$, where the ground state lies in the uniform $\pi$-flux sector. The Dirac nodes in the phase diagram are topologically stable. At the critical (red) line $J'_x/J'_y=J_x/J_y$, fermions form a Fermi surface (FS). The other two critical (blue) lines are defined as $J'_x=J_5^2/(4J_x)$ or $J'_y=J_5^2/(4J_y)$. 
}
\label{PhaseD}
\end{figure}

When $0<J_5\ll \Delta_v$, (so the ground state lies in the $\pi$-flux sector,)   
it is clear that $E_{+,\bk}$ is always gapped. The conditions for $E_{-,\bk}$ to have gapless excitations are:
\begin{eqnarray}
&&J_xJ'_x \cos^2 k_x+J_yJ'_y\sin^2 k_y =J_5^2/4,\\
&&(J_xJ'_y-J_yJ'_x)\cos k_x \sin k_y =0. 
\end{eqnarray} 
For simplicity, we consider the case in which $J_x,J_y \gg J_5 >0$  
 so that $J_5\ll \Delta_v$ is 
  satisfied
 for arbitrary $J'_x$ and $J'_y$. 
From the two 
conditions we obtain the phase diagram shown in Fig. \ref{PhaseD} as a function of  $J'_x$ and $J'_y$: (i) When $J'_x>J_5^2/(4J_x)$, $J'_y>J_5^2/(4J_y)$, and $J'_x/J'_y\neq J_x/J_y$, the  
fermion spectrum has eight Dirac nodes at $(\pm \pi/2\pm\theta_x,0)$ and $(\pm \pi/2,\pm  \theta_y)$ with $\theta_\lambda = \arcsin (J_5/\sqrt{4J_\lambda J'_\lambda})$; (ii) When $J'_x>J_5^2/(4J_x)$ and $J'_y<J_5^2/(4J_x)$, there are four Dirac nodes at $(\pm \pi/2\pm\theta_x,0)$; (iii) When $J'_x<J_5^2/(4J_x)$ and $J'_y>J_5^2/(4J_x)$, there are also four Dirac nodes but at $(\pm \pi/2,\pm\theta_y)$; (iv) When $J'_x<J_5^2/(4J_x)$ and $J'_y<J_5^2/(4J_y)$, all fermionic excitations are gapped.

For $0<J_5\ll \Delta_v$, the Dirac nodes, if they exist, are topologically stable in the sense of Wigner-Von Neumann theorem. Any weak translational and time reversal invariant perturbation  
only shifts the positions of the nodes \cite{HY}. 
Consequently, the present phase with Dirac nodes is characteristic of a stable quantum phase of matter, {\it i.e.}
 an algebraic spin liquid \cite{Wen01,Hermele04,Hermele05}.

It is worth noting that along the critical line $J'_x/J'_y=J_x/J_y$ and $J'_x>J_5^2/(4J_x)$, corresponding to the red line in Fig. \ref{PhaseD}, the discrete nodes broaden into a line of nodes. In short, in this special case the present model realizes a spinon Fermi surface. Since spin-3/2 operators can be written as a bosonic bilinear in term of two Schwinger bosons, $S^\alpha=b^\dag_s \sigma^\alpha_{ss'}b_{s'}/2$, $s=\uparrow,\downarrow$ with the constraint $b^\dag_\uparrow b_\uparrow +b^\dag_\downarrow b_\downarrow=3$, the present spin model can be written as a bosonic model, albeit one with  four-body interactions. It follows as a corollary 
 that by tuning a single coupling constant to a critical value, a purely bosonic model can exhibit an emergent Fermi surface. 

Upon approach to the critical line $J'_x=J_5^2/(4J_x)$ from the eight node phase, both of the two Dirac cones at $\pm ( \pi/2+\theta_x,0)$ approach $(\pi,0)$ leading, at criticality, to a single node with the unusual dispersion $E_\bk\approx\sqrt{Aq_x^4+B q_y^2})$ for small $|\bq|=|\bk-(\pi,0)|$, where $A$ and $B$ are constants depending on $J'_\lambda$, $J_\lambda$, and $J_5$. Another two nodes at $\pm (\pi/2-\theta_x,0)$ approach $(0,0)$ leading to the same unusual dispersion. Similar physics is obtained at the other critical line $J'_y=J_5^2/(4J_y)$. 

{\it Large-$J_5$ limit}: 
The ground-states for $1/J_5=0$ form
the low energy manifold $S^z_i=\pm 3/2$ for each $i$. 
Defining an effective spin-1/2 $\vec \sigma$, 
such that $S^z_i=\pm 3/2$ corresponds to $\sigma^z_i=\pm 1$, and employing degenerate perturbation theory in $J_\lambda$ and $J'_\lambda$, the  
 effective Hamiltonian can be shown be $H_\eff=J_\eff\sum_i \sigma^x_i\sigma^y_{i+\hat x}\sigma^x_{i+\hat x+\hat y}\sigma^y_{i+\hat y}$, $J_\eff=(J_x-J'_x)^2(J_y-J'_y)^2/(16J_5^3)$, which is  exactly Wen's  
 plaquette model \cite{Wen03}. In terms of fluxes, $H_\eff=\sum_i J_\eff\exp[i\phi_i]$. Consequently, 
 the ground state is still in the $\pi$-flux sector.   
We expect that the gapped phase in small $J_5$ limit can be adiabatically connected to the large $J_5$ gapped phase.

{\it Discussion:}
It is straightforward to generalize the GMM to other lattices in 2D and also to higher dimensions. For the 2D triangular lattice, a quantum spin-7/2 (or spin-$\frac{1}{2}$-$\frac{1}{2}$-$\frac{1}{2}$ \cite{Wen03PRD}) model can be defined via the seven Gamma matrices. Due to the non-bipartiteness of the lattice, the model spontaneously breaks TRS. Thus, it is expected that a chiral spin liquid \cite{Yao07} could be realized in this model on a triangular lattice \cite{HY}. 

Acknowledgments: We would like to thank Eduardo Fradkin, Zhengcheng Gu, Taylor Hughes, Alexei Kitaev, Robert B. Laughlin, Joseph Maciejko, Joel Moore, Chetan Nayak, Xiaoliang Qi, and Shivaji Sondhi for helpful discussions. We are grateful to Xiao-Gang Wen for pointing out 
a different interpretation of the present Gamma matrix model in Ref. \cite{Wen03PRD}. S.A.K. and H.Y. are supported in part by DOE grant DEFG03-01ER45925. S.C.Z. is supported by NSF DMR-0342832 and DOE grant DE-AC03-76SF00515. 
H.Y. thanks the support by a Stanford Graduate Fellowship.


\begin{thebibliography}{30}

\bibitem{Anderson87} P.W. Anderson, Science {\bf 235}, 1196 (1987).

\bibitem{wenleereview}  For a review, see P. A. Lee, N. Nagaosa, X.-G. Wen, \rmp{78}{17}{2006}.

\bibitem{Moessner01} R. Moessner and S. Sondhi, \prl{86}{1881}{2001}.

\bibitem{Rokhsar88} D. S. Rokhsar and S. A. Kivelson, Phys. Rev. Lett. {\bf 61}, 2376 (1988).

\bibitem{Anderson73} P. W. Anderson, Mater. Res. Bull. {\bf 8}, 153 (1973).

\bibitem{Kivelson87} S. A. Kivelson, D. S. Rokhsar, and J. P. Sethna, Phys. Rev. B {\bf 35}, 8865 (1987).

\bibitem{Wen03} X. G. Wen, \prl{90}{016803}{2003}.

\bibitem{Shimizu03} Y. Shimizu {\it et al}, Phys. Rev. Lett. {\bf 91}, 107001 (2003).

\bibitem{Yamashita08} S. Yamashita {\it et al}, Nat. Phys. {\bf 4}, 459 (2008).

\bibitem{Sachdev08} Y. Qi, C. Xu, and S. Sachdev, Arxiv: 0809.0694.

\bibitem{Helton07} J. S. Helton {\it et al}, \prl{98}{107204}{2007}.

\bibitem{Kitaev06} A. Kitaev, Annals of Physics {\bf 321}, 2 (2006).

\bibitem{Zhang04} S. Murakami, N. Nagaosa, and S. C. Zhang, \prb{69}{235206}{2004}.

\bibitem{Wen03PRD} X.-G. Wen, Phys. Rev. D, {\bf 68}, 065003 (2003).

\bibitem{Lieb94} E. H. Lieb, \prl{73}{2158}{1994}.

\bibitem{Baskaran07} G. Baskaran, S. Mandal, and R. Shankar, \prl{98}{247201}{2007}.

\bibitem{Chen07} H.-D.Chen and J.-P.Hu, \prb{76}{193101}{2007}.

\bibitem{Wen01}  W. Rantner and X.-G. Wen, \prl{86}{3871}{2001}.

\bibitem{Hermele04} M. Hermele {\it et al},
 \prb{70}{214437}{2004}.

\bibitem{Hermele05} M. Hermele, T. Senthil, and M. P. A. Fisher, \prb{72}{104404}{2005}.

\bibitem{Shimizu06} Y. Shimizu {\it et al}, Phys. Rev. B {\bf 73}, 140407(R) (2006).

\bibitem{Yao07} H. Yao and S. A. Kivelson, \prl{99}{247203}{2007}

\bibitem{HY} H. Yao {\it et al}, unpublished. 

\end{thebibliography}
\end{document}